\documentclass[preprint]{elsarticle}
\usepackage{amsmath}
\usepackage{graphicx}
\usepackage{amssymb}
\usepackage{epstopdf}
\usepackage[colorlinks=true]{hyperref}
\DeclareGraphicsExtensions{.pdf,.eps}
\begin{document}

\title{Symbolic dynamics and synchronization of coupled map networks with multiple delays}

\author{Fatihcan M. Atay}

\ead{atay@member.ams.org} \address{Max Planck Institute
for Mathematics in the Sciences, Inselstr.~22, 04103 Leipzig, Germany}


\author{Sarika Jalan}

\ead{physarik@nus.edu.sg} \address{Department of Physics and Centre for Computational Science and Engineering, 
National University of Singapore, 117456, Republic of Singapore}

\author{J{\"u}rgen Jost}

\ead{jjost@mis.mpg.de} \address{Max Planck Institute for
Mathematics in the Sciences, Inselstr.~22, 04103 Leipzig, Germany}
\address{Santa Fe Institute, 1399 Hyde Park Road, Santa Fe, NM 87501, USA.}

\date{}

\begin{abstract}
We use symbolic dynamics to study discrete-time dynamical systems
with multiple time delays. We exploit the concept of avoiding sets, which arise
from specific non-generating partitions of the phase space and restrict
the occurrence of certain symbol sequences related to the characteristics
of the dynamics. In particular, we show that the resulting forbidden sequences
are closely related to the time delays in the system. We present 
two applications to coupled map lattices, namely (1) detecting synchronization
and (2) determining unknown values of the transmission delays in
networks with possibly directed and weighted connections and measurement
noise. The method is applicable to multi-dimensional as well as set-valued
maps, and to networks with time-varying delays and connection structure.
\end{abstract}


\maketitle

\noindent \textit{Preprint}. For final version, see:
 \textit{Physics Letters A} \textbf{375} (2010) 130--135.   \\
\href{http://dx.doi.org/10.1016/j.physleta.2010.10.044}{doi: 10.1016/j.physleta.2010.10.044}

\section{Introduction}

Symbolic dynamics is a versatile tool for describing the complicated
time evolution of dynamical systems, the Smale horseshoe being a famous
prototype \cite{smale}. Here, instead of representing a trajectory
by a continuum of numbers, one watches the alternation of symbols from a finite alphabet.
In the process some information is ``lost" but certain
important invariants and robust properties of the dynamics may be
kept \cite{book-sym2,book-sym2b}. Most studies of symbolic dynamics are based
on the so-called generating partition \cite{generating-book} of the
phase space, for which topological entropy achieves its maximum \cite{topological-entropy}.
Symbolic dynamics based on generating partitions plays a crucial role
in understanding many different properties of dynamical systems. However,
finding generating partitions is generally a difficult problem 
\cite{generating-eg,generating-eg2,generating-eg3,generating-eg4,generating-eg5}.
Some consequences of using misplaced partitions have been investigated
in \cite{mispart}. Nevertheless, certain non-generating partitions
have recently been shown to have particular uses. Specifically, appropriately
chosen partitions that restrict the appearance of certain symbolic
subsequences have been used for distinguishing random from deterministic
time series \cite{complexity06}, and for investigating the collective
behavior of coupled systems \cite{chaos06}. 

On the other hand, time delays arise naturally in the modeling of many physical systems.
In spatially extended systems, such as networks, delays are a consequence of the fact
that signals cannot be transmitted instantly over distances. An additional source of delays can be the time it takes for
each unit to process the information it receives before it acts on
it. Interestingly, networks of dynamical systems can still synchronize their actions under certain conditions despite time delays, although the synchronized solution can be very different from an undelayed network \cite{delay,complexity04}. 
Sometimes the value of the delays are unknown or may be changing in time, and the determination of the delay value is a problem in itself 
\cite{Lepri93,Buenner98,So2002,Rad2003,Drakunov2006,Yu2008,Yu2009}.
In studying the collective behavior of networks of dynamical systems,
it is therefore both realistic and important to take time delays into
account in the modeling and to develop techniques to handle the subsequent
complications in the analysis.

In the following, we use symbolic dynamics for the study of discrete-time
systems with multiple connection delays. Although significant time delays are common in physical and biological systems, the effects of delays on the symbolic
dynamics have not received much attention so far. We extend the notion
of ``forbidden words", that is, symbol sequences whose appearance is restricted by the dynamical constraints, to systems
with delays. The basic idea is to derive forbidden words for the
delayed system from the properties of the undelayed map. 
We show how forbidden sequences are related to the time delays
in the system and how this information provides useful information
about the dynamics. We apply the theoretical findings to two important
practical problems: Detecting synchrony in a large network with multiple delays using measurements from only a few nodes, and determining unknown values of the delays in the network. As might be expected from the
``crudeness'' introduced by symbolic dynamics, the method has
a certain robustness against noise.

\section{Symbolic dynamics for delayed maps}

\label{sec:map} Let $f:S\rightarrow S$ be a map on a subset $S$ of $\mathbb{R}^{n}$,
and consider the dynamical system defined by the iteration rule 
\begin{equation}
	x(t+1)=f(x(t)),
		\label{system}
\end{equation}
 where the iteration step $t\in\mathbb{Z}$ plays the role of discrete
time. Let $\{S_{i}:i=1,\ldots,m\}$ be a partition of $S$, i.e.,
a collection of nonempty and mutually disjoint subsets satisfying
$\cup_{i=1}^{m}S_{i}=S$. (We assume $m>1$ to prevent trivial cases.)
The symbolic dynamics corresponding to (\ref{system}) is the sequence
of symbols $\{{\dots,s_{t-1},s_{t},s_{t+1},\dots\}}$, where $s_{t}=i$
if $x(t)\in S_{i}$. 
In the usual grammatical analogy, the symbols $\{1,2,\dots,m\}$
form the \textit{alphabet}, and finite symbol sequences 
are called \textit{words}. We say the set $S_{i}$ \emph{avoids} $S_{j}$
under $f$ if \begin{equation}
f(S_{i})\cap S_{j}=\emptyset.\label{av}\end{equation}
 Clearly, if $S_{i}$ avoids $S_{j}$, so does any of its subsets.
We also refer to a \emph{self-avoiding set} if (\ref{av}) holds with
$i=j$. The significance of avoiding sets is that they yield forbidden
words: If $S_{i}$ avoids $S_{j}$, then the symbolic dynamics for
(\ref{system}) cannot contain the symbol sequence $ij$. The notion
is extended in a straightforward way to the $k$th iterate of $f$.
Thus, if $f^{k}(S_{i})\cap S_{j}=\emptyset$, then the symbol sequence
for the dynamics cannot contain any subsequence of the form 
$i\,(*\cdots *)_{(k-1)}\, j$, where
$(*\cdots *)_{(k-1)}$ denotes $k-1$ arbitrary symbols.). 
In other words, a symbol block of length $k+1$ that starts with $i$ cannot end with $j$. 
This constrains the symbol sequences
that can be generated by a given map, and provides a robust method
to distinguish between different systems by inspecting their symbolic
dynamics. As examples of avoiding sets, we mention that, for the familiar
unimodal maps of the interval $[0,1]$, such as the tent or logistic
maps, the set $(x^{\ast},1]$ and its subsets are self-avoiding, where
$x^{\ast}$ denotes the positive fixed point of $f$.

More generally, partitions that contain avoiding sets can always be
found. We give a constructive proof. Suppose one starts with some
partition $\{S_{1},\dots,S_{m}\}$ of $m$ sets for which (\ref{av})
does not hold for any $i,j$; that is, \begin{equation}
f(S_{i})\cap S_{j}\neq\emptyset\quad\forall i,j.\label{qq31}\end{equation}
Now fix some pair $(i,j)$, $i\neq j$. 
Partition the set $S_{i}$ further into two disjoint sets as $S_{i}=S_{i}^{1}\cup S_{i}^{2}$,
where \begin{align*}
S_{i}^{1} & :=f^{-1}(S_{j})\cap S_{i},\\
S_{i}^{2} & :=S_{i}\backslash S_{i}^{1}.\end{align*}
Thus, $S_{i}^{1}$ and $S_{i}^{2}$ contain those points of $S_{i}$
that are mapped to $S_{j}$ and those that are not mapped to $S_{j}$,
respectively, by the function $f$. Note that by definition $f(S_{i}^{2})\cap S_{j}=\emptyset$,
that is, $S_{i}^{2}$ avoids $S_{j}$ under $f$. Furthermore, $S_{i}^{1}\neq\emptyset$
by (\ref{qq31}), and $S_{i}^{2}\neq\emptyset$ because otherwise
we would have $f(S_{i})\subset S_{j}$, which would imply $f(S_{i})\cap S_{i}=\emptyset$
(since $S_{i}$ and $S_{j}$ are disjoint sets by assumption), which
would contradict (\ref{qq31}). Hence, we can define a new partition
of $m+1$ nonempty and mutually disjoint sets\begin{equation}
\{S_{1},\dots,S_{i-1},S_{i}^{1},S_{i+1},\dots,S_{m},S_{i}^{2}\},\label{qq32}\end{equation}
which is obtained from the original one by replacing $S_{i}$ by $S_{i}^{1}$
and adding $S_{i}^{2}$ , in which the set $S_{i}^{2}$ avoids $S_{j}$.
The same argument can be used to construct self-avoiding sets: Assume
$f(S_{i})\not\subset S_{i}$ (otherwise further partition $S_{i}$
to obtain a set which is not invariant under $f$, 
which is  possible except for the trivial case when $f$ is the identity
map.) Define $S_{i}^{1}=f^{-1}(S_{i})\cap S_{i}$ and $S_{i}^{2}=S_{i}\backslash S_{i}^{1}$.
Then $S_{i}^{2}$ is a self-avoiding set in the new partition (\ref{qq32}).
Hence, it is possible to modify a given partition so that the sequence
$ij$ (or $ii$) never occurs in the symbolic dynamics.

The above arguments apply equally well to discrete-time inclusions
\begin{equation}
x(t+1) \in F(x(t))\label{eq:inclusion}
\end{equation}
where $F$ is a set-valued function in $\mathbb{R}^{n}$. This
case arises, e.g., when the actual function $f$ is not precisely
known or is constructed from data, or when measurements are contaminated
with noise, so the value $f(x)$ can only be determined up to some
error bound. For instance, the point-value $f(x)$ plus the ``error
disc'' could be used to define the set-value $F(x)$. We define
avoiding sets for set-valued functions $F$ in the same way
through (\ref{av}), which similarly yield forbidden sequences for
(\ref{eq:inclusion}). Hence, all results we present here remain valid
when equalities are replaced by set inclusions.

To apply the above ideas to delayed dynamics, we first consider the
following extension of (\ref{system}), 
\begin{equation}
x(t+1)=(1-\varepsilon)f(x(t))+\varepsilon f(x(t-\tau)),
\label{delayed}
\end{equation}
where $\tau\in\mathbb{Z}^{+}$ is the time delay and $\varepsilon\in[0,1]$
is a parameter measuring the relative weight of the past in determining
the next state. The significance of Eq.~(\ref{delayed}) is that
it governs the behavior of the synchronous solutions of coupled map
networks with transmission delays \cite{delay,complexity04}, which are studied
in Section \ref{sec:network}. The domain $S$ of the map $f$ is
required to be a convex set in order for the iterations (\ref{delayed})
to be meaningful. Clearly, for $\tau=0$ (\ref{delayed}) reduces
to (\ref{system}). The symbolic dynamics is defined as before, but
we define avoiding sets slightly differently. We say the set $S_{i}$
\emph{convexly avoids} $S_{j}$ under $f$ if $\mbox{conv}(f(S_{i}))\cap S_{j}=\emptyset$,
where {}``conv'' denotes the convex hull of a set. Such sets give
rise to forbidden sequences as follows: If $S_{i}$ convexly avoids
$S_{j}$ under $f$, then the symbolic sequence 
\begin{equation}
i\,\underbrace{(*\dots*)}_{\tau-1}\, i\, j
\label{forbidden}
\end{equation}
 is not possible for the delayed system (\ref{delayed}). 
This is a consequence of (\ref{delayed})
and the observation that if $x(t)$ and $x(t-\tau)$ are both in $S_{i}$,
then any convex combination of $f(x(t))$ and $f(x(t-\tau))$ belongs
to the convex hull of $f(S_{i})$ and so lies outside of $S_{j}$.
Similarly, if $S_{i}$ is convexly self-avoiding, then any sequence
of $\tau+1$ symbols that begin and end with $i$ cannot be followed
by another $i$. An important observation is that, although the dynamics
of (\ref{delayed}) can vary greatly with $\varepsilon$ \cite{delay},
the forbidden sequences (\ref{forbidden}) are independent of the
value of $\varepsilon$. Thus, (\ref{forbidden}) remains a forbidden sequence for the symbolic dynamics of the time-dependent equation 
\begin{equation}
x(t+1)=(1-\varepsilon(t))f(x(t))+\varepsilon(t) f(x(t-\tau)),
\end{equation}
where $\varepsilon:\mathbb{Z} \to [0,1]$  is allowed to be 
a function of time.

Finally, we generalize to equations with multiple delays of the form
\begin{equation}
x(t+1)=\sum_{\tau=0}^{\tau_{\max}}\varepsilon_{\tau}f(x(t-\tau)),\label{delayed2}\end{equation}
where the coefficients $\varepsilon_{\tau}$ are nonnegative and satisfy
$\sum_{\tau=0}^{\tau_{\max}}\varepsilon_{\tau}=1$. Such equations
govern the synchronous solutions of coupled maps with multiple delays,
as will be shown in Section \ref{sec:multiple}. Note that the right
hand side of (\ref{delayed2}) lies in the convex hull of the set
$\{f(x(t-\tau)):\tau=0,\dots,\tau_{\max}\}$. Therefore, if $S_{i}$
convexly avoids $S_{j}$ under $f$, then the sequence 
\begin{equation}
\underbrace{i\, i\dots i}_{\tau_{\max}+1}\, j\label{forbidden2}
\end{equation}
is forbidden for (\ref{delayed2}); that is, a sequence of consecutive
$i$'s of length $\tau_{\max}+1$ cannot be followed by a $j$. Again,
this result is independent of the values of the coefficients $\varepsilon_{\tau}$, so the latter can be allowed to vary with time, subject to the constraint that they remain nonnegative and sum up to 1.
Further restrictions are obtained if some $\varepsilon_{m}$ is identically zero,
in which case (\ref{forbidden2}) will be forbidden even when the
symbol at position $(\tau_{\max}+1-m)$ is replaced by an arbitrary
symbol in the alphabet. 

The additional condition of convexity of the sets in case of the delayed
dynamics does not present an extra restriction in many practical situations.
In fact, one often measures a single component, say the first one,
of the $n$-dimensional vector $x=(x_{1},\dots,x_{n})$. In this case,
a simple partition of $S$ given by the disjoint union $S=S_{1}\cup S_{2},$
where 
\[
S_{1}=\{(x_{1},\dots,x_{n})\in\mathbb{R}^{n}:x_{1}<x^{\ast}\}
\]
 and $x^{\ast}$ is a scalar threshold value, which can be chosen
to make both $S_{1}$ and $S_{2}$ nonempty. It is easy to see that
both $S_{1}$ and $S_{2}$ defined in this way are convex whenever
$S$ is convex. Such partitions are almost surely non-generating,
so the corresponding symbol sequences do not capture all features
of the dynamics. (For a discussion of obtaining partitions in a simple
setting, see \cite[Section VII]{chaos06}.) Nevertheless, it will
be seen that they still contain important information that can be
utilized to study some important aspects about the delayed dynamics.

\section{Coupled map networks with time delay}

\label{sec:network}We now move from single maps to networks of coupled
maps, in the context of a model which is sometimes referred to as
the coupled map lattice \cite{CML}. We consider a general form allowing
arbitrary coupling topology, directed and weighted connections, as
well as time delay along the connections: 
\begin{equation}
x_{i}(t+1)=f(x_{i}(t))+\frac{\varepsilon}{k_{i}}\sum_{j=1}^{N}a_{ij}\left[f(x_{j}(t-\tau))-f(x_{i}(t))\right].
\label{cml}
\end{equation}
Here $x_{i}(t)$ is the state of the $i$th unit at time $t$, $i=1,\dots,N$,
$a_{ij}\ge 0$ is the weight on the link from $j$ to $i$ (zero if
there is no link), $\varepsilon\in[0,1]$ is the coupling strength,
and $k_{i}=\sum_{j}a_{ij}$ is the weighted 
\textit{in-degree} of node $i$. (It is understood that the summation
term is set to zero in (\ref{cml}) for any unit for which $k_{i}$
is zero.) The delay $\tau$ is the time it takes for the information
from a unit to reach its neighbors and be processed. The system is
said to synchronize if $|x_{i}(t)-x_{j}(t)|\rightarrow0$ as $t\rightarrow\infty$
for all $i,j$ and all initial conditions from some open set. In this
case, the state of every node asymptotically approaches the same \textit{synchronous
solution} $x(t)$, whose dynamics is governed by (\ref{system}) and
(\ref{delayed}), respectively, depending on whether the delay $\tau$
is zero or nonzero. In the absence of delays, various aspects of the
network have been studied using symbolic dynamics \cite{chaos06,Pethel2006}.
Our focus here is on the delayed case. 

It is known that the network (\ref{cml}) can synchronize even under
delays, where the units are unaware of the present states of their
neighbors but still can act in unison \cite{delay}. The important
distinction from the undelayed case, however, is that the synchronous
dynamics $x(t)$ is no longer identical to the isolated dynamics (\ref{system})
of the units, but is governed by the delayed equation (\ref{delayed}).
A consequence is that the overall system (\ref{cml}) can exhibit
a much wider range of behavior than its constituent units through
the coordination of their actions 
\cite{complexity04}. An important problem is to determine whether a large
network is synchronized using information from just a few nodes. As
a first application, we study this problem in delayed networks.

\begin{figure*}
\centering 
\includegraphics[scale=0.7]{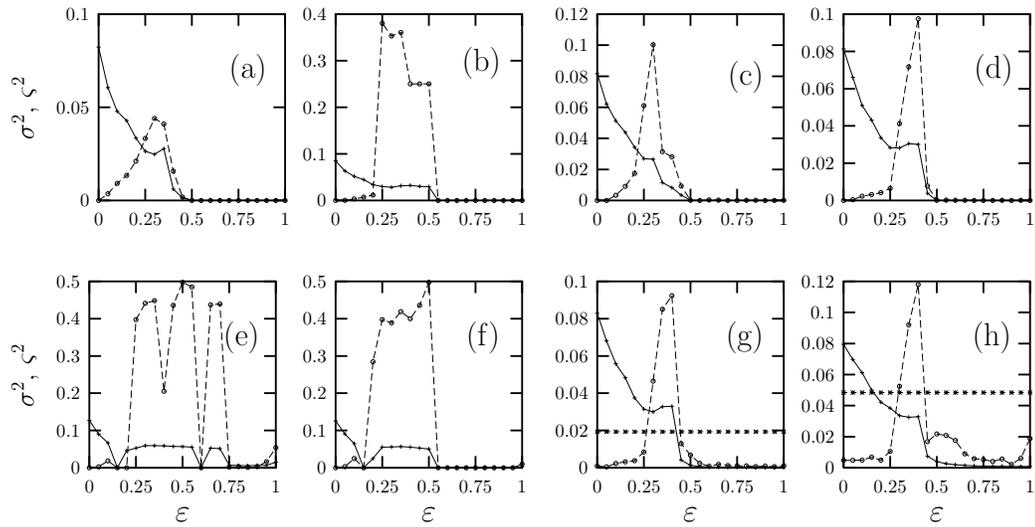}  
\caption{Detecting network synchrony by comparing the transition probabilities
from measurements at a single node with those of the synchronous solution
(\ref{delayed}). Average deviations $\varsigma^{2}$ in transition
probabilities (dashed line) and synchronization measure $\sigma^{2}$
(solid line) are plotted against the coupling strength $\varepsilon$.
The synchronous regime where $\sigma^{2}=0$ coincides with the regions
where $\varsigma^{2}=0$. The plots (a)--(d) are for a globally coupled
network of 20 nodes, and delay values (a) $\tau=0$, (b) $\tau=1$,
(c) $\tau=2$, (d) $\tau=3$. (e) and (f) are plotted, respectively,
for a random network of 100 nodes and a scale-free network of 200
nodes, where $\tau=3$ for both. Subfigures (g) and (h) are for the
same network and delay as in (d), but with additive Gaussian noise
of $2\%$ and $5\%$, respectively, in the measurements.}
\label{fig1n} 
\end{figure*}

Normally the symbol sequences observed from a node of a network can
vary widely between the nodes. However, in the synchronized state
$x_{i}(t)=x(t)$ for all $i$, so that the symbolic sequences observed
from a node will be subject to the same constraints as that generated
by (\ref{delayed}). 
The choice of the node is arbitrary so long as the network is capable
of chaotic synchronization (which is the case for the choice of parameters in our example systems).
This gives a method of detecting synchronization of the network by
choosing an arbitrary node and calculating the transition probabilities
of symbol subsequences: From the relative frequencies of occurrence
of subsequences of the form (\ref{forbidden}) in the measured time
series, one estimates the transition probabilities $P(j|(i*\dots*i)_{\tau+1})$,
that is, the conditional probability that a sequence of length $\tau+1$
starting and ending with $i$ is followed by $j$. Letting $\varsigma^{2}$
denote the average squared difference between the observed transition
probabilities of the network and those of (\ref{delayed}), synchronization
is signaled when $\varsigma^{2}=0$.

Fig.~\ref{fig1n} illustrates the relation between synchronization
and forbidden sequences, 
for the chaotic tent map $f(x)=1-2|x-\tfrac{1}{2}|$ and the partition
\begin{equation}
S_{1}=[0,x^{\ast}],\quad S_{2}=(x^{\ast},1],\label{partition}\end{equation}
where $x^{\ast}=2/3$ is the fixed point of $f$. Note that $S_{2}$
is a (convexly) self-avoiding set under $f$. We evolve (\ref{cml})
starting from random initial conditions and estimate the transition
probabilities using time series of length 1000 from a randomly selected
node. (We note that the length of the time series used is independent
of the network size.) 
Synchronization occurs when the variance 
$\sigma^{2}=\left\langle \frac{1}{N-1}\sum_{i}[x_{i}(t)-\bar{x}(t)]^{2}\right\rangle _{t}$
drops to zero, where $\bar{x}(t)=\frac{1}{N}\sum_{i}x_{i}(t)$ denotes
the average over the nodes of the network and $\left\langle \dots\right\rangle _{t}$
denotes an average over time. As seen from Fig.~\ref{fig1n}, the
region for synchronization exactly coincides with the region where
the transition probabilities for the network are identical to those
of Eq.~(\ref{delayed}). Hence, regardless of network topology and
size, both synchronized and unsynchronized behavior can be detected
over the whole range of coupling strengths using only measurements
from an arbitrarily selected node. Moreover, Figure \ref{fig1n}(g-h)
show that the {}``crudeness'' introduced by using symbolic sequences
also provides some robustness against noise. 

\begin{figure}
\centering 
\includegraphics[width=0.6\columnwidth]{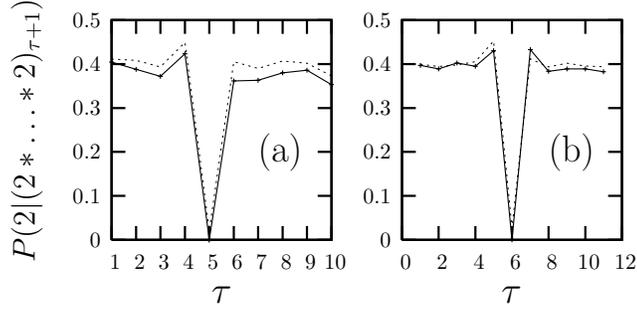}
\caption{Finding the unknown value of delay in a synchronized network using
measurements from an arbitrarily selected node. The observed probability
$P(2|(2*\dots*2)_{\tau+1})$ is plotted against $\tau$. The true
value of $\tau$ corresponds to the point where the probability drops
to zero. The network has 20 nodes that are globally coupled with $\varepsilon=0.75$,
where the true value of the delay is (a) 5 and (b) 6. The dotted lines
are in the presence of $10\%$ noise.}
\label{fig2n} 
\end{figure}

As a second application, we consider the reverse problem of determining
the value of the delay $\tau$ in (\ref{delayed}) from observed symbolic
dynamics. For this purpose, we check the presence of subsequences
of the form (\ref{forbidden}) of various lengths, knowing that such
a sequence of length $\tau+2$ would be forbidden. Plotting the occurrence
frequencies of (\ref{forbidden}) against $\tau$, the actual value
of the delay is found at the point where the frequency drops to zero
(or attains its minimum, in the presence of small noise). Similarly,
the value of the delay in the network (\ref{cml}) can be found from
a knowledge of its synchrony. A practical situation is when the value
of $\tau$ is unknown but the network is known to be synchronized
or can be made to synchronize by the adjustment of control parameters.
The value of $\tau$ can then be obtained by using the measurements
from a node and checking the presence of the forbidden sequences (\ref{forbidden}).
Fig.~\ref{fig2n} gives an illustration for the tent map and the
partition (\ref{partition}), by plotting the probability $P(2|(2*\dots*2)_{\tau+1})$
versus $\tau$, that is, the conditional probability that a sequence
of length $\tau+1$ starting and ending with 2 is followed by another
2. By the arguments above, such a sequence cannot occur for the synchronized
dynamics (\ref{delayed}) since $S_{2}$ is convexly self-avoiding.
The true value of the delay is thus found at the point where the occurrence
probability of the sequence drops to zero. Fig.~\ref{fig2n} shows
that the method works well also under noise.

\section{Time-varying delay and connection structure}

The arguments above remain valid also when the connection topology is changing with time, as in the network
\begin{equation*}
x_{i}(t+1)=f(x_{i}(t))+\frac{\varepsilon(t)}{k_{i}(t)}\sum_{j=1}^{N}a_{ij}(t)\left[f(x_{j}(t-\tau))-f(x_{i}(t))\right],
\end{equation*}
where $\varepsilon(t) \in [0,1]$ and $k_{i}(t)=\sum_{j}a_{ij}(t)$ for all $t\in \mathbb{Z} $.
This is a consequence of the 
observation that the synchronized solution does not depend on the network topology and its forbidden sequences (\ref{forbidden}) are independent of $\varepsilon$. The conditions for synchronization, of course, depend on the connection structure. 
In the undelayed case, synchronization conditions involve the existence of spanning trees of the union graphs and can be quantified in terms of the Hajnal diameter of infinite sequences of connection matrices \cite{Lu,Lu2}. On the other hand, the precise conditions for synchronization of delayed time-varying networks is a more involved problem.

\begin{figure}[tb]
	\centering
	\includegraphics[scale=0.5]{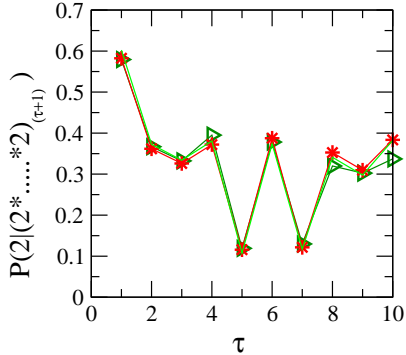}
	\caption{Finding the unknown values of time-varying delay in a synchronized network using 
measurements from an arbitrarily selected node. The delay randomly takes one of the values $\{5,7\}$ with equal probability at each time step. The observed probability
$P(2|(2*\dots*2)_{\tau+1})$ is plotted against $\tau$, displaying marked dips at the true values of the delay. The network consists of 20 all-to-all coupled nodes, with coupling strengths of $\varepsilon=0.85$, $0.86$ and $0.9$, shown by the three curves.}
	\label{fig:time-var-delay}
\end{figure}

A further generalization is to allow delays that change with time; $\tau = \tau(t)$. In this case, sequences such as (\ref{forbidden}) will be forbidden at some time point $t$ for the corresponding value of $\tau(t)$. For instance, if the partition set $S_i$ convexly avoids $S_j$ under $f$, then $s_{t+1}\neq j$ whenever $s_t = s_{t-\tau(t)}=i$. Even the precise time dependence  $\tau(t)$ is not known, one can still obtain useful information by studying such subsequences. Thus,  
if $\tau(t)$ often assumes some value $k$, then the $(k+1)$-symbol block  
\begin{equation}
i\,\underbrace{(*\dots*)}_{k-1}\, i\, j
\label{forbidden-k}
\end{equation}
will correspondingly appear more seldom; hence, rather than being forbidden within the whole symbolic history, it will have reduced frequency of occurrence. This observation helps determine the unknown values of the time-varying delay. 
To illustrate, we return to our example of coupled tent maps used for Figs.~\ref{fig1n} and \ref{fig2n}, this time considering a time-varying connection delay $\tau$ whose value at each time step is chosen randomly from the set $\{5,7\}$. In Fig.~\ref{fig:time-var-delay} we plot the occurrence frequencies of the sequences (\ref{forbidden}) for various values of $\tau$. 
In contrast to Fig.~\ref{fig2n}, the frequency does not drop to zero but displays two marked dips at the values $\tau=5$ and $\tau=7$, agreeing with the fact that the delay was randomly switching between 5 and 7.

\section{Multiple delays}

\label{sec:multiple} The foregoing ideas can be extended to systems
with multiple delays, e.g., to the coupled map network 
\begin{equation}
x_{i}(t+1)=f(x_{i}(t))+\frac{\varepsilon}{k_{i}}\sum_{j=1}^{N}a_{ij}\left[f(x_{j}(t-\tau_{ij}))-f(x_{i}(t))\right],\label{cml2}
\end{equation}
where $\tau_{ij}$ denotes the transmission delay from $j$ to $i$.
Whereas the network (\ref{cml}) with a constant delay always admits
a synchronous solution, one needs additional conditions in the case
(\ref{cml2}) of multiple delays. Namely, Eq.~(\ref{cml2}) has non-constant
synchronized solutions provided that the fraction of weighted incoming
connections having a given value of delay is the same for every vertex.
To see this, suppose $x_{i}(t)=x(t)$ for all $i$. Substituting into
(\ref{cml2}) and rearranging, we have 
\begin{equation}
x(t+1)-(1-\varepsilon)f(x(t)=\varepsilon\sum_{j=1}^{N}\frac{a_{ij}}{k_{i}}f(x(t-\tau_{ij})),\label{cml-multi}
\end{equation}
where we have used the fact that $k_{i}=\sum_{j}a_{ij}$. One can
decompose the summation further over the delay values since the delays
$\tau_{ij}$ are integers, thus obtaining \begin{equation}
x(t+1)-(1-\varepsilon)f(x(t))=\varepsilon\sum_{\tau=0}^{\tau_{\max}}f(x(t-\tau))\sum_{j\in J_{i}(\tau)}\frac{a_{ij}}{k_{i}},\label{qq21}\end{equation}
where $\tau_{\max}=\max_{i,j}\{\tau_{ij}\}$ is the maximum delay
in the network and $J_{i}(\tau)=\{j:\tau_{ij}=\tau\}$ is the index
set of the connections to $i$ that are subject to a delay of precisely
$\tau$. Now if $f$ is constant over the synchronous trajectory $\mathcal{X}:=\{x(t):t\in\mathbb{Z}\}$,
then the term $f(x(t-\tau$)) can be taken outside the summation and
the double summation adds up to 1, reducing the equation to (\ref{delayed}).
This case happens, in particular, when the synchronous solution is
constant. In general, however, for non-constant synchronous solutions,
$f$ will not be constant over $\mathcal{X}$. In this case, since
the left hand side of (\ref{qq21}) is independent of $i$, we require
that the quantity \[
\sum_{j\in J_{i}(\tau)}\frac{a_{ij}}{k_{i}}\]
be also independent of $i$. In other words, for any given value of
delay, the weighted fraction of incoming links having that delay value
should be the same for each node. We let $p_{\tau}=\sum_{j\in J_{i}(\tau)}a_{ij}/k_{i}$
denote this common fraction, and define\[
\varepsilon_{\tau}=\left\{ \begin{array}{lc}
\varepsilon p_{\tau}, & \mbox{if }\tau\ge1\\
(1-\varepsilon)+\varepsilon p_{0}, & \mbox{if }\tau=0\end{array}\right..\]
Note that $\sum_{\tau}p_{\tau}=1$; therefore, $\sum_{\tau}\varepsilon_{\tau}=1$.Thus (\ref{qq21}) becomes
\begin{equation*}
x(t+1)=\sum_{\tau=0}^{\tau_{\max}}\varepsilon_{\tau}f(x(t-\tau)),
\end{equation*}
which is the same as Eq.~(\ref{delayed2}) considered in Section \ref{sec:map}.
Thus the synchronous solution $x(t)$ of the system (\ref{cml-multi})
with multiple delays obeys (\ref{delayed2}), and 
by the results of Section \ref{sec:map}, symbolic sequences of the form 
(\ref{forbidden2}) are forbidden for the synchronous dynamics.
Such symbol sequences can thus be used
to determine whether a delay value of $m$ is present in the network
(\ref{cml2}), yielding a systematic way of finding the values of
all delays from a knowledge of synchrony. Conversely, synchronization
can be detected by comparing the transition probabilities of symbolic
sequences from an arbitrary node to those of (\ref{delayed2}) if
the delays are known.

\begin{figure}
\begin{centering}
\includegraphics[width=0.45\columnwidth]{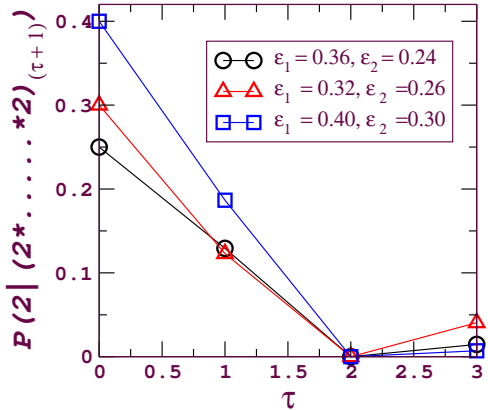}  
\par\end{centering}

\caption{The value of the largest delay in networks with multiple delays is
obtained from the non-occurrence of ($\tau_{\max}+1$) consecutive
symbols corresponding to a self-avoiding set.\label{fig:multiple-delay}}

\end{figure}

As an example of networks with multiple delays, we consider a network
of four nodes arranged on a circle, where each node is coupled to
its nearest neighbors on its left and right with delay equal to 1
and to its far neighbor on the opposite side with delay equal to 2.
The local map is the chaotic shift map $f(x)=2x$ (mod 1) on the unit
interval, which we partition as $S_{1}=[0,0.25)$, $S_{2}=[0.25,0.5)$,
$S_{3}=[0.5,0.75)$, and $S_{4}=[0.75,1]$. It is easy to see that
both $S_{2}$ and $S_{3}$ are convexly self-avoiding under $f$.
Hence, from (\ref{forbidden2}), symbol sequences containing $\tau_{\max}+2$
consecutive 2's are forbidden. One can then check the lengths of uninterrupted
subsequences of 2's (or 3's) from a node of the synchronized network
and determine the largest value of the delay. Fig.~\ref{fig:multiple-delay}
shows that subsequences of four consecutive 2's are not observed,
implying that $\tau_{\max}$ is indeed equal to 2. Furthermore, the
result is independent of the value of the coupling strength.

\section{Discussion and conclusion}

In this paper we have used symbolic dynamics to study discrete-time
systems and networks with time delays. 
We have derived forbidden symbol
sequences for the delayed system from the properties of the undelayed
map. Although the partitions used are usually non-generating, the
forbidden sequences are related to certain characteristics of the
dynamics, and in particular to delays. Consequently, the value of
the delay in the system can be determined by the presence and absence
of such sequences. Conversely, a knowledge of the delays enables one
to detect synchronization (or phase locking) in the network
using measurements from a single node. 
The method has the advantage of being based on a phase-space
partition that is much easier to obtain than a generating partition.
Furthermore, it can utilize rather short measurements from a single
node of the network. The computations are therefore fast and independent
of the network size, and do not require knowledge of the connection
structure. As such, they can complement or be an alternative to existing
techniques for detecting synchronization based on phase-space reconstruction
\cite{predict,tseries,tseries2} and methods for estimating delay times 
\cite{Lepri93,Buenner98,So2002,Rad2003,Drakunov2006,Yu2008,Yu2009}.

Although we have restricted our discussion to complete synchronization, 
the ideas apply also to some other types of collective behavior, for instance to \emph{phase-locked solutions} and \emph{traveling waves}. In such collective states, the symbol sequences of all nodes in the network are identical except for a time shift (that depends on the particular node), which does not change symbol statistics.
Hence, the forbidden sequences can be derived as before from the properties of the local map, with the same implications as in the applications presented here. 

Beyond the coupled map lattice model (\ref{cml}), there is also a
growing interest in more general coupling schemes such as \begin{equation}
x_{i}(t+1)=f(x_{i}(t))+\frac{\varepsilon}{k_{i}}\sum_{j=1}^{N}a_{ij}g(x_{i}(t),x_{j}(t-\tau)).\label{general}\end{equation}
Synchronized solutions $s(t)$ of (\ref{general}) satisfy
\begin{equation}
s(t+1)=h(s(t),s(t-\tau))\label{sync-general}
\end{equation}
where the function $h$ is defined by
\begin{equation}
h(x,y):=f(x)+\varepsilon g(x,y).
\label{h}
\end{equation}
Thus, for instance (\ref{delayed}) becomes a special case with $h(x,y)=(1-\varepsilon)f(x)+\varepsilon f(y)$.
The notion of avoiding sets can be extended to the more general case:
We say that the set $S_{i}$ avoids $S_{j}$ under $h$ if
\begin{equation}
h(S_{i},S_{i})\cap S_{j}=\emptyset.
\label{av-general}
\end{equation}
If (\ref{av-general}) holds, then the symbol sequence (\ref{forbidden})
is a forbidden sequence for the dynamics (\ref{sync-general}). 
More generally, if $h(S_{i},S_{k})\cap S_{j}=\emptyset$
for some sets in the partition, then the symbol sequence\[
k\,\underbrace{(*\dots*)}_{\tau-1}\, i\, j\]
will be forbidden. With a knowledge of forbidden sequences, one can
pursue the line of reasoning of the previous sections to derive results
about the
coupled system (\ref{general}). The additional challenge now is
to relate the avoiding sets to the properties of the functions $f$ and $g$
and the coupling coefficient $\varepsilon$. 
The difficulty is of course not unique to the symbolic-dynamics
approach, since the dynamics of the system (\ref{general}),
with more parameters in its structure,
is not easy to characterize in its full generality, although there has been some recent progress in this direction. For example,
for the undelayed case, conditions for the stability of the synchronous state have been given in \cite{Nonlinearity09}, and Ref.~\cite{EPL10} has shown the range of rich dynamics such systems can exhibit at synchrony. For the delayed case, however, considerably less is known at present.  

Finally, we note that our treatment is based on systems with known dynamics, such as Eq.~(\ref{system}), or models of approximately known dynamics, such as the discrete-time inclusion (\ref{eq:inclusion}). 
On the other hand, in certain important applications only a time series of measurements is available without any detailed knowledge of the dynamical process generating it. 
In the absence of \emph{a priori} information about the forbidden sequences, the applicability of the methods of this paper is restricted. 
Nevertheless, it may still be possible to exploit similar ideas in combination with the methods of time series analysis. One possibility is to use \emph{a posteriori} statistics of subsequences from the time series. 
In fact, here one need not confine himself to forbidden sequences but instead can use statistical information of symbol sequences to compare the network's behavior with that of individual units; see e.g. \cite{chaos06} for an example in the undelayed case. 
Another alternative is to first build a mathematical model of the dynamical process from time series using several well-established methods \cite{tseries}. Once a model is fit to data, the analysis presented here can be carried out as before.

\end{document}